# Demonstration of a genetic-algorithm-optimized cavity-based waveguide crossing

PENGFEI XU,[1] YANFENG ZHANG,[1,*] ZENGKAI SHAO,[1] YUJIE CHEN,[1] AND SIYUAN YU[1,2]

[1]*State Key Laboratory of Optoelectronic Materials and Technologies, School of Electronics and Information Technology, Sun Yat-sen University, Guangzhou 510275, China*

[2]*Photonics Group, Merchant Venturers School of Engineering, University of Bristol, Bristol BS8 1UB, UK*

*\*zhangyf33@mail.sysu.edu.cn*

**Abstract:** A compact, single-etch silicon photonics waveguide crossing based on low quality factor cavity and optimized by a genetic algorithm is demonstrated. The device has 0.1~0.3 dB insertion loss and < -35 dB crosstalk per crossing for fundamental TE mode in the 1550-1600 nm wavelength band, with a footprint of 5×5 μm$^2$.

## 1. Introduction

Waveguide crossing is a fundamental element in photonic integrated circuits (PICs). High performance crossings with low insertion loss, low crosstalk, low backscattering, and small footprint are particularly challenging to achieve in silicon photonics due to the high refractive index contrast. In the past two decades, much literature about silicon waveguide crossings or intersections has been published. Three-dimensional overpass [1] can provide very high performance but is complicated to fabricate. Single layer crossing schemes, though are more convenient to fabricate, require elaborate designs to achieve high performance, as summarized in Fig. 1 and Table 1. These can be divided in to two main types － the cavity-based waveguide crossings and multi-mode waveguide-based (MMW-based) crossings.

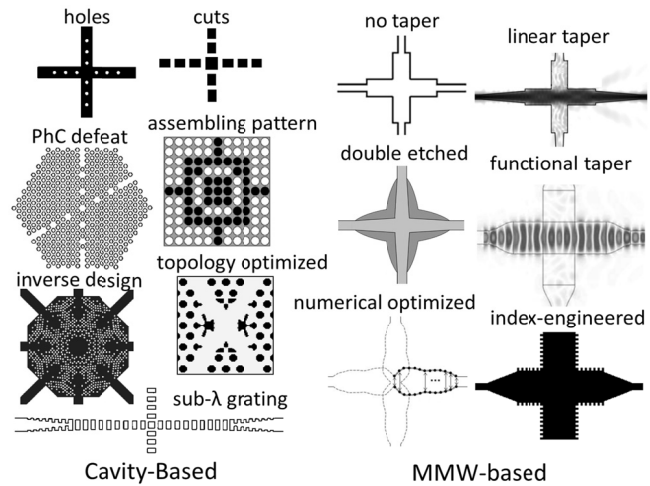

Fig. 1. Waveguide crossing published in recent years, which can roughly summarized into two categories: cavity-based crossing and multi-mode waveguide based crossing.

Many high performance waveguide crossings based on multi-mode waveguide have been reported [2,3,4], among which a genetic-algorithm-optimized waveguide crossing [5] and Bloch-mode-based crossings [6,7] have demonstrated very low insertion loss of < 0.05dB and low crosstalk of < -40 dB. Their footprints are about 9×9 μm$^2$. A MMW-based waveguide crossing with reduced footprint of about 6 μm$^2$ [8-10] provides slightly higher insertion loss of 0.1~0.3 dB with narrower bandwidth.

Cavity-based waveguide crossings reported earlier [11,12] had slightly higher insertion loss and more severe wavelength dependency. They have been mainly studied in photonic crystals waveguides [13-15]. A waveguide crossing designed with the sub-wavelength grating (SWG) [16] supported both TE and TM modes at an extremely low insertion loss of about 0.02 dB. However the SWG crossing required a taper with relative high loss of ~0.3 dB to transit between a plain waveguide and the SWG waveguide crossing, the resulting total insertion loss was ~0.6 dB.

Table 1. Summaries of published representative single-etched waveguide crossings.

| Type | Year | Insertion loss | crosstalk | bandwidth | footprint | references |
|---|---|---|---|---|---|---|
| Cavity | 1998 | 0.2 dB | -30 dB | — | 5×5 μm$^2$ | [11] |
| Cavity | 1999 | 0.17 dB | -30 dB | — | 1.2×1.2 μm$^2$ | [12] |
| MMW | 2006 | 0.2 dB | -35 dB | 100 nm | 10×10 μm$^2$ | [3] |
| MMW | 2009 | 0.1 dB | -40 dB | 20 nm | 6×6 μm$^2$ | [8] |
| MMW | 2010 | 0.3 dB | -40 dB | 100 nm | 5×5 μm$^2$ | [9] |
| Cavity | 2010 | 0.6 dB | -40 dB | 60 nm | 9×9 μm$^2$ | [16] |
| MMW | 2013 | 0.1-0.3 dB | -38 dB | 80 nm | 6×6 μm$^2$ | [10] |
| MMW | 2013 | 0.02 dB | -40 dB | 90 nm | 8×8 μm$^2$ | [6] |
| MMW | 2013 | 0.02 dB | -37 dB | 100 nm | 9×9 μm$^2$ | [5] |
| MMW | 2014 | 0.04 dB | -35 dB | 100 nm | 9×9 μm$^2$ | [7] |
| Cavity | 2017 | 0.75 dB | -22.5 dB | 60 nm | 5×5 μm$^2$ | [17] |
| Cavity | — | 0.1~0.3 dB | -35 dB | 50 nm | 5×5 μm$^2$ | This work |

In recent years, with the improvement of computer performance, many novel photonic device designs have been produced by numerical optimization algorithms [17]. Numerically-optimized photonic devices often demonstrate superior features comparing to conventional devices, e.g., smaller footprint, lower loss, wider spectrum bandwidth and customized functions.

Genetic algorithm (GA) is a widely-used numerical algorithm in device optimization. In a dynamic design model, parameters are defined to regulate the geometric structure or the material properties. All the independent parameters are ordered and collected in an array to form a "gene", i.e. a parameter value set, and GA is applied to tentatively adjust the parameter values in the gene one by one. Once the gene is changed, the model simulation result as a feedback will evaluate the gene revision, and then decide to reserve or discard the revision. After sufficient number of iterations, when any gene revision can no longer increase the model performance, an optimal or quasi-optimal device design is considered to be achieved.

Examples of GA-optimized photonic devices include a silicon nitride-on-SOI grating coupler with low-fabrication complexity [18], an ultra-small footprint polarization rotator [19], and a broadband optical waveguide coupler [20]. An ultra-compact power splitter with a QR code-like nanostructure is designed and demonstrated [21]. These examples verified the advantages of GA-optimization as a valid design tool in photonics. In this work, we aim to use GA optimization to create a cavity-based waveguide crossing with low-insertion loss, low crosstalk, and small footprint as well.

## 2. Device design

The waveguide crossing, as illustrated in Fig. 2 (a), is designed for 220-nm thick and 500-nm wide silicon waveguides. The design aims at a compact footprint of 5×5 μm$^2$. The crossing comprises a distinctive sub-wavelength structure that forms a low-Q optical cavity. The resonance within the sub-wavelength structure cavity ensures the transmission from one port to the opposite port with low insertion loss; while the crosstalk to other ports and back-reflection can be suppressed. Different from the conventional cavity, the waveguide crossing cavity is designed with a GA-based square pattern with a scale of 51×51, i.e. 2601 pixels at a pixel size of 100 × 100 nm$^2$. Due to the 4-fold symmetry, here we only show one half-quadrant with 351 independent pixels in Fig. 2(b-c). The design process is illustrated in Fig. 2(b-f). The device is designed by changing the square pattern by means of GA optimization, performed to tentatively assemble these independent pixels aimed at reducing the insertion loss and improve the channel isolation at 1550 nm wavelength.

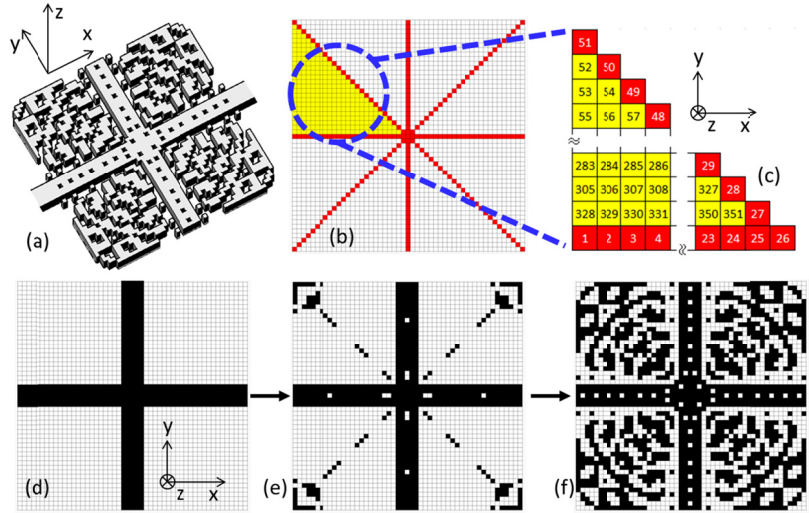

Fig. 2. (a) The proposed cavity-based waveguide crossing. (b) Pattern framework with a pixel size about $100 \times 100$ nm$^2$ and only one half-quadrant are independent pixels due to the 4-fold symmetry. (c) The independent pixels with ordered indices, which forms a "gene", i.e. binary value array. (d) Initial pattern: plain waveguide crossing. (e) One of the mid-optimization patterns. (f) Final optimized pattern.

The pattern optimization starts from a plain waveguide crossing shown in Fig. 2(d). The pixels are filled with either black (silicon preserved) or white (silicon etched away). The binary value array (black or white) of the independent pixels listed in Fig. 2(c) forms a 'gene'. The GA sequentially and tentatively flips the binary value in the "gene" array to the opposite value. When the gene is revised, the simulated insertion loss and crosstalk at 1550 nm are used as a feedback to evaluate the performance of the new "gene". If the insertion loss is reduced, the new "gene" is adapted, otherwise the revision is abandoned. The process is repeated until any flip on any pixel can no longer reduce the insertion loss, the optimization will be achieved. A mid-optimization pattern is shown in Fig. 2(e), and a final optimized pattern is shown in Fig. 2(f). Note that, in addition to the structures surrounding the waveguides, the optimization process created holes inside the waveguides similar to photonic crystal cavities.

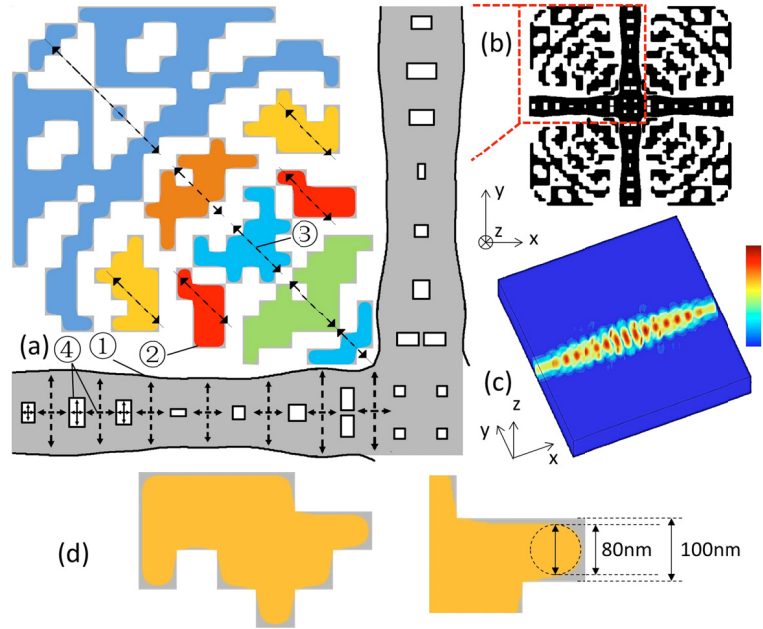

Fig. 3. (a) An additional multi-wavelength optimization process, including ① waveguide width optimization, ② sharp corner removing, ③ colored structures position tuning, and ④ holes size and position tuning. (b) The optimized device layout. (c) The 3D profile of $|\mathbf{E}|^2$ for the device with a fundamental TE mode input from one port at 1550 nm wavelength, where $\mathbf{E}$ is the electric field component of the optical wave studied. (d) Sharp corners are rounded by graphical edge feathering with a radius of 10 nm.

In the above process, the GA optimizes the crossing insertion loss and crosstalk only at 1550 nm wavelength. Therefore, the crossing performance can be sensitive to wavelength detuning, i.e. the optical spectrum bandwidth of the crossing can be limited. In addition, the 100×100 nm$^2$ pixels are difficult to realize in fabrication because their sharp corners require complex proximity correction in the Electron Beam Lithography (EBL) process. Therefore, an additional step of fabrication-friendly GA optimization is applied to improve the fabrication tolerance and multi-wavelength performance, as shown in Fig. 3(a). Firstly, the width of the waveguide in the crossing region is modulated by dividing the waveguide into 7 sections. The width of the 7 sections at the positions indicated by the vertical double arrows is varied and the shape of the waveguide in between is determined by spline interpolation, as illustrated in ① of Fig. 3(a). Secondly, the sharp corners are rounded by graphical edge feathering with a radius of 10 nm , as shown in ② of Fig. 3(a) and (d). Thirdly, the positions of the structures outside the waveguide are also fine-tuned, with the 7 colored blocks outside the waveguide slightly moved diagonally relative to the device symmetry center, as shown in ③ of Fig. 3(a). Lastly, the width, length, and position of all 8 holes in each waveguide (a total of 24 independent parameters) are also optimized by GA to enhance multi-wavelength performance in the 1550~1600 nm range, as shown in ④ of Fig. 3(a). In these additional optimization steps, a total of 38 new structural parameters (7 widths, 7 diagonal positions of the colored blocks, and the 24 hole parameters) are fine-tuned by GA with a step size of 1 nm. The final device layout is shown in Fig. 3(b).

The Lumerical® finite-difference-time-domain (FDTD) simulation package is used to verify the performance of the design. The three-dimensional intensity distribution is plotted in Fig 3(c). The wavelength responses of the insertion loss, back-reflection and crosstalk are also examined in Lumerical® FDTD, of which results are plotted in Fig. 4. The insertion loss of the crossing in the 1550~1600 nm wavelength region is predicted to be 0.05~0.2 dB, the crosstalk to be < -35 dB, and the reflection lower than -20 dB.

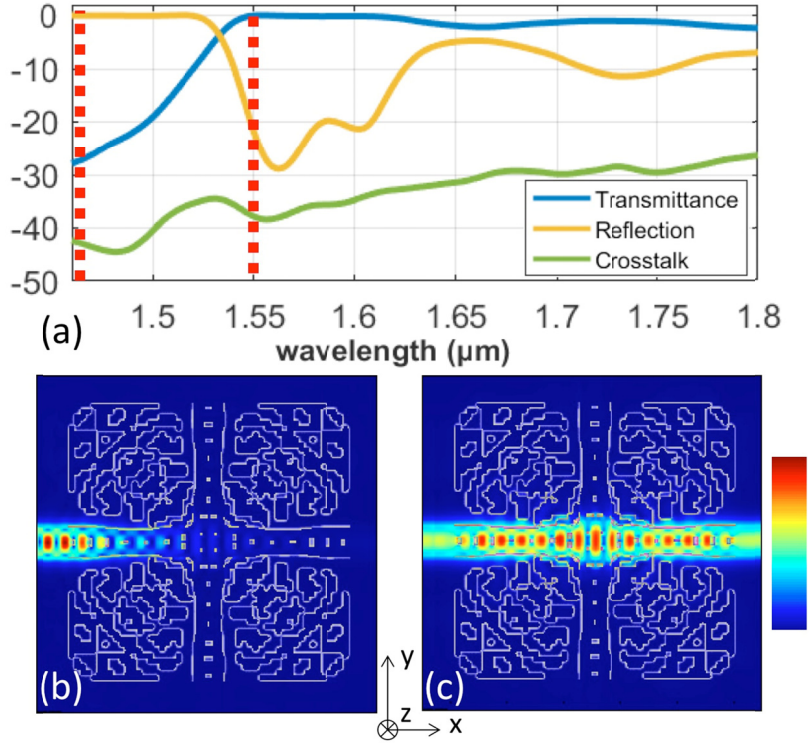

Fig. 4. (a) The Lumerical FDTD simulated wavelength response of the transmittance, crosstalk, and reflection. The corresponding electric field distributions at 1450nm and 1550nm wavelength are shown in (b) and (c), respectively.

The crossing is predicted to behave like a photonic crystal (PhC). When the wavelength is located in the 'forbidden' wavelength region resulting in total reflection and the propagation will be blocked due to the destructive interference shown in Fig.4(b), whereas within the crossing wavelength window, due to constructive interference, light can pass the intersection with low insertion loss shown in Fig. 4(c). In this design, a 30-dB transmission suppression ratio is achieved at 1450 nm, without increasing the crosstalk level.

## 3. Device fabrication and test

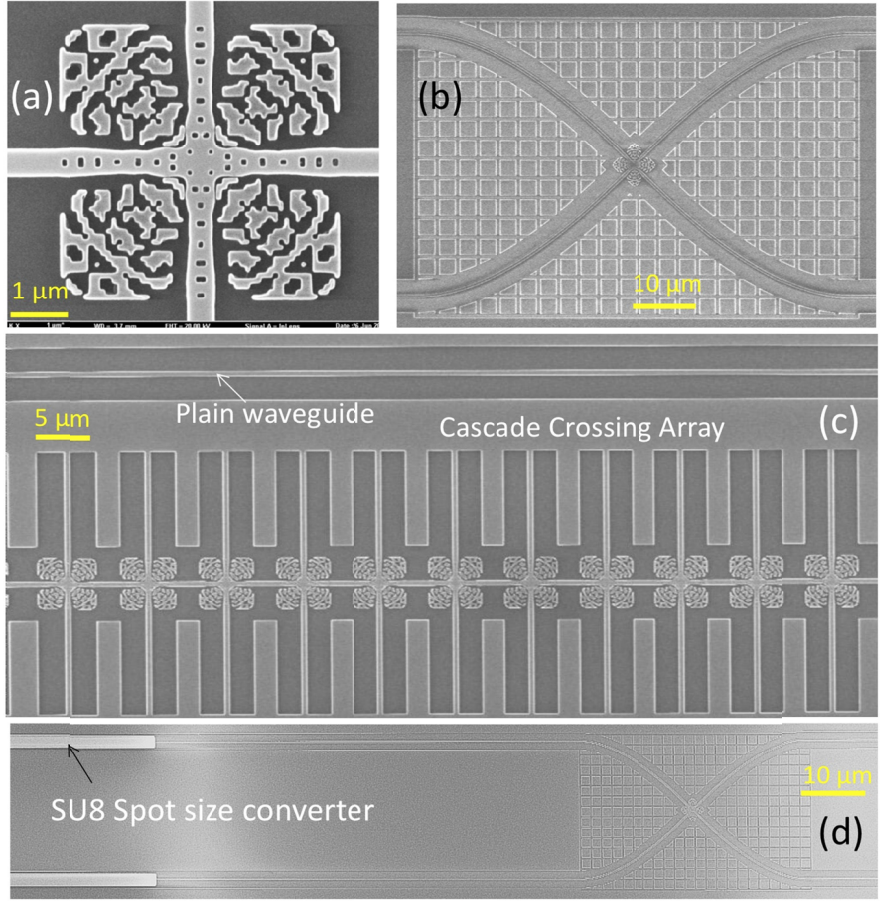

Fig. 5. (a) The zoom-in SEM image of the waveguide crossing. (b) SEM image of the waveguide intersection for measuring the crosstalk. (c) Cascaded waveguide crossing for precise insertion loss measurement. (d) SEM image of the SU8 spot-size converter.

The fabrication of the waveguide crossing is patterned by Electron Beam Lithography (EBL) and Inductively Coupled Plasma (ICP) etching on an SOI wafer with a 220-nm thick silicon layer. All structures are etched together in one step. The fabricated device examined under Scanning Electron Microscope (SEM) is shown in Fig. 3(a). In order to measure the insertion loss precisely, a long series of waveguide crossing is fabricated as shown in Fig. 5(c). A two waveguide intersection, shown in Fig 5(b), was also fabricated on the same chip for measuring the crosstalk. The grid around the waveguides and waveguide crossing is etched to block optical leakage which may transmit through the substrate. A spot-size converter with a 400-μm long inverted silicon taper and a SU8 polymer encapsulating waveguide of 3.5×3.5 μm$^2$ cross-section is fabricated to reduce the cleaved facet coupling loss, shown in Fig. 5(d). Reference silicon waveguides without waveguide crossing are also fabricated on the same wafer for comparison and calibration. The measured transmittance and crosstalk of the crossing devices are calibrated against the reference waveguides which has the same spot-size converter facet coupler.

The experimental measurement is carried out using a Keysight® 8164B lightwave measurement system, and the results are shown in Fig. 6. In order to measure the insertion loss precisely, cascaded crossings as exemplified in Fig.5(c) with 1, 5, 10, and 20 crossings are measured, respectively. The insertion loss curves are shown in Fig. 6 (a). Due to the wavelength dependency of the PhC-like low-Q cavity-based waveguide crossing, the shorter wavelength region can be well filtered. Especially in cascaded crossing arrays, the wavelength cut-off will be more obvious. However, in the transmitted wavelength region of 1550-1600 nm, the insertion loss is small, as plotted and fitted in Fig. 6(b). The averaged insertion loss is estimated by linear fitting with value between 0.1-0.3 dB per crossing, plotted in Fig. 6(c). The crosstalk level is measured using the waveguide intersection of Fig. 5(b), where the waveguide ports are marked. The $|S21|^2$ and $|S34|^2$ crosstalk are measured to be -35 dB or lower across the wavelength region, as plotted in Fig. 6(d).

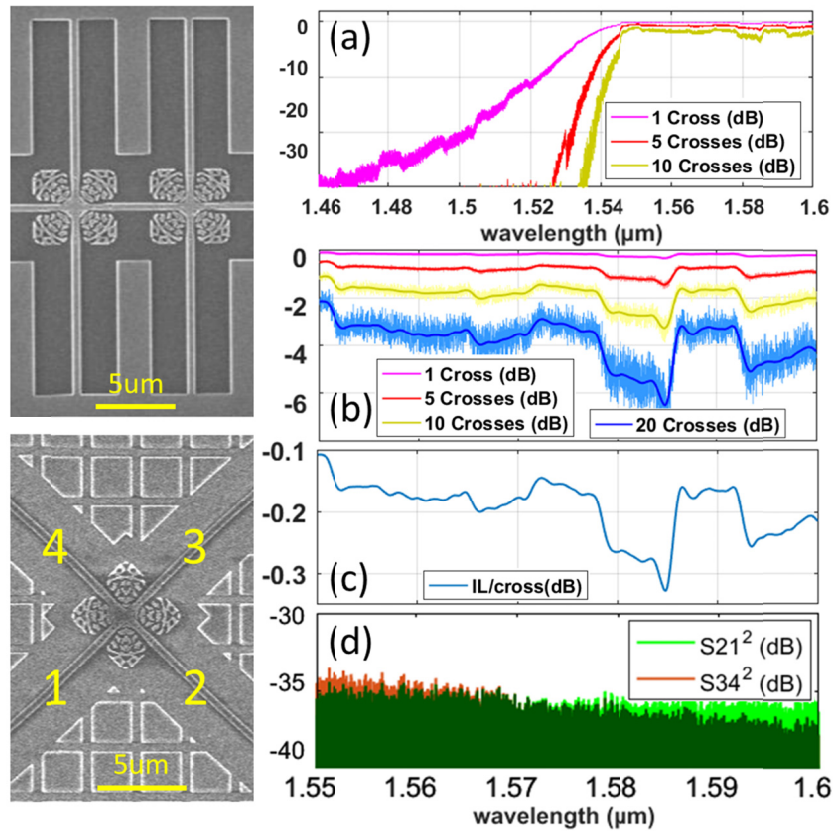

Fig. 6. (a) Measured insertion loss wavelength responses with different numbers of the cascaded waveguide crossing at 1460-1600 nm wavelengths. (b) The zoomed insertion loss in 1550-1600 nm. (c) Calculated average insertion loss per crossing. (d) Measured crosstalk level with respect to corresponding ports.

## 4. Conclusion

A GA-optimized compact waveguide crossing for TE mode is designed, fabricated, and tested. A low insertion loss of about 0.1~0.3 dB with a low crosstalk of < -35 dB per crossing is obtained. The demonstrated performance of the waveguide crossing is acceptable to many practical applications and the compact design and small footprint make it competitive in a limited chip space for many photonic-based applications.

## Funding

This work is supported by National Key Research and Development Program of China (2016YFB0402503), National Basic Research Program of China (973 Program) (2014CB340000), National Natural Science Foundations of China (11774437, 61323001, 11690031, 61490715, 51403244), Natural Science Foundation of Guangdong Province (2014A030313104), Science and Technology Program of Guangzhou (201707020017), and Fundamental Research Funds for the Central Universities of China (Sun Yat-sen University: 17lgzd06, 16lgjc16, 15lgpy04, 15lgzs095, 15lgjc25).

## References

1. A. M. Jones, C. T. DeRose, A. L. Lentine, D. C. Trotter, A. L. Starbuck, and R. A. Norwood, "Ultra-low crosstalk, CMOS compatible waveguide crossings for densely integrated photonic interconnection networks," Opt. Express 21, 12002-12013 (2013).
2. H. Liu, H. Tam, P. K. A. Wai, and E. Pun, "Low-loss waveguide crossing using a multimode interference structure," Optics Communications 241, 99-104 (2004).
3. H. Chen, and A. W. Poon, "Low-loss multimode-interference-based crossings for silicon wire waveguides," IEEE photonics technology letters 18, 2260-2262 (2006).
4. W. Bogaerts, P. Dumon, D. V. Thourhout, and R. Baets, "Low-loss, low-cross-talk crossings for silicon-on-insulator nanophotonic waveguides," Opt. Lett. 32, 2801-2803 (2007).

5. Y. Ma, Y. Zhang, S. Yang, A. Novack, R. Ding, A. E.-J. Lim, G.-Q. Lo, T. Baehr-Jones, and M. Hochberg, "Ultralow loss single layer submicron silicon waveguide crossing for SOI optical interconnect," Opt. Express 21, 29374-29382 (2013).

6. Y. Zhang, A. Hosseini, X. Xu, D. Kwong, and R. T. Chen, "Ultralow-loss silicon waveguide crossing using Bloch modes in index-engineered cascaded multimode-interference couplers," Opt. Lett. 38, 3608-3611 (2013).

7. Y. Liu, J. M. Shainline, X. Zeng, and M. A. Popović, "Ultra-low-loss CMOS-compatible waveguide crossing arrays based on multimode Bloch waves and imaginary coupling," Opt. Lett. 39, 335-338 (2014).

8. P. Sanchis, P. Villalba, F. Cuesta, A. Håkansson, A. Griol, J. V. Galán, A. Brimont, and J. Martí, "Highly efficient crossing structure for silicon-on-insulator waveguides," Opt. Lett. 34, 2760-2762 (2009).

9. C.-H. Chen, and C.-H. Chiu, "Taper-integrated multimode-interference based waveguide crossing design," IEEE Journal of Quantum Electronics 46, 1656-1661 (2010).

10. Y. Zhang, S. Yang, A. E.-J. Lim, G.-Q. Lo, C. Galland, T. Baehr-Jones, and M. Hochberg, "A CMOS-compatible, low-loss, and low-crosstalk silicon waveguide crossing," IEEE Photon. Technol. Lett 25, 422-425 (2013).

11. S. G. Johnson, C. Manolatou, S. Fan, P. R. Villeneuve, J. D. Joannopoulos, and H. A. Haus, "Elimination of cross talk in waveguide intersections," Opt. Lett. 23, 1855-1857 (1998).

12. C. Manolatou, S. G. Johnson, S. Fan, P. R. Villeneuve, H. A. Haus, and J. D. Joannopoulos, "High-density integrated optics," Journal of Lightwave Technology 17, 1682-1692 (1999).

13. S. Lan, and H. Ishikawa, "Broadband waveguide intersections with low cross talk in photonic crystal circuits," Opt. Lett. 27, 1567-1569 (2002).

14. J. Yang, S. F. Mingaleev, M. Schillinger, D. A. B. Miller, F. Shanhui, and K. Busch, "Wannier basis design and optimization of a photonic crystal waveguide crossing," IEEE Photonics Technology Letters 17, 1875-1877 (2005).

15. Y. Yu, M. Heuck, S. Ek, N. Kuznetsova, K. Yvind, and J. Mørk, "Experimental demonstration of a four-port photonic crystal cross-waveguide structure," Applied Physics Letters 101, 251113 (2012).

16. P. J. Bock, P. Cheben, J. H. Schmid, J. Lapointe, A. Delâge, D.-X. Xu, S. Janz, A. Densmore, and T. J. Hall, "Subwavelength grating crossings for silicon wire waveguides," Opt. Express 18, 16146-16155 (2010).

17. L. Lu, M. Zhang, F. Zhou, W. Chang, J. Tang, D. Li, X. Ren, Z. Pan, M. Cheng, and D. Liu, "Inverse-designed ultra-compact star-crossings based on PhC-like subwavelength structures for optical intercross connect," Opt. Express 25, 18355-18364 (2017).

18. P. Xu, Y. Zhang, Z. Shao, L. Liu, L. Zhou, C. Yang, Y. Chen, and S. Yu, "High-efficiency wideband SiNx-on-SOI grating coupler with low fabrication complexity," Opt. Lett. 42, 3391-3394 (2017).

19. Z. Yu, H. Cui, and X. Sun, "Genetic-algorithm-optimized wideband on-chip polarization rotator with an ultrasmall footprint," Opt. Lett. 42, 3093-3096 (2017).

20. P. H. Fu, Y. C. Tu, and D. W. Huang, "Broadband optical waveguide couplers with arbitrary coupling ratios designed using a genetic algorithm," Opt. Express 24, 30547-30561 (2016).

21. K. Xu, L. Liu, X. Wen, W. Sun, N. Zhang, N. Yi, S. Sun, S. Xiao, and Q. Song, "Integrated photonic power divider with arbitrary power ratios," Opt. Lett. 42, 855-858 (2017)